\documentclass[english]{article}
\usepackage{mathptmx}

\usepackage[T1]{fontenc}
\usepackage[utf8]{luainputenc}
\usepackage[a4paper]{geometry}
\usepackage{babel}
\usepackage{float}
\usepackage{amsmath}
\usepackage{graphicx}
\usepackage{esint}
\usepackage[unicode=true,pdfusetitle,
 bookmarks=true,bookmarksnumbered=false,bookmarksopen=false,
 breaklinks=false,pdfborder={0 0 1},backref=false,colorlinks=false]
 {hyperref}

\makeatletter

\providecommand{\tabularnewline}{\\}

\usepackage{babel}

\makeatother

\begin{document}

\title{Color octet neutrino as the source of the IceCube $PeV$ energy neutrino
events}

\author{A. N. Akay$^{1}$, U. Kaya$^{1}$, S. Sultansoy$^{1,2}$ \\
\textit{\small $^{1}$TOBB Univesity of Economics and Technology,
Ankara, Turkey}\\
\textit{\small $^{2}$ANAS Institute of Physics, Baku, Azerbaijan}}
\maketitle
\begin{abstract}
\textit{It is shown that two $PeV$ events observed by the IceCube
collaboration can be interpreted as resonance production of color
octet neutrinos.}
\end{abstract}
$\,$

Recent observation of two $PeV$ energy events by the IceCube collaboration
\cite{M. G. Aarten et al.} may be and indication of new physics at
$TeV$ scale. A number of candidates for origin of these events have
been considered, for example, resonance production of leptoquarks
\cite{V. Barger}, $PeV$-scale decaying dark matter \cite{A. Esmaili,B. Feldstein}
and so on. In this letter we argue that observed events can be interpreted
as the first manifestation of color octet neutrinos. Corresponding
Feynman diagram is presented in Fig. 1.

\begin{figure}[H]
\begin{centering}
\includegraphics[scale=0.5]{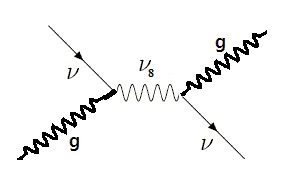}
\par\end{centering}

\caption{Resonant $\nu_{8}$ production in the ultra high energy neutrino nucleon
scattering.}
\end{figure}

Color octet neutrinos and leptons, as well as color sextet (decuplet
etc) quarks are predicted by preonic models with colored preons {[}5-8{]}.
Recently, color octet leptons have been came to forefront again {[}9-13{]}.
It should be noted that concerning preon models color octet neutrinos
has the same status as color octet leptons, and the status of both
of them is similar to exited leptons and neutrinos, which are widely
investigated in HEP experiments.

There are two strong arguments favoring preon models: inflation of
“fundamental” particles and free parameters in the SM (other BSM models,
including SUSY, drastically increase the number of free parameters)
and mixing of “fundamental” quarks and leptons. The first one, namely
“inflation”, historically results in discovery of new level of matter
two times during the last century: periodical table of chemical elements
was clarified by Rutherford experiment, inflation of hadrons results
in quark model (see Table 1 from {[}14{]}). 

According to PDG \cite{PDG} current exclusion limit for color octet
neutrino is $M_{\nu_{8}}>110$$GeV$ assuming $\nu_{8}\rightarrow\nu+g$
decay. This value is obtained from Tevatron data, rough estimations
show that $M_{\nu_{8}}$ below $400\, GeV$ could be excluded by current
LHC data.

The interaction lagrangian for color octet neutrinos is given by 

$\,$

$L=\frac{1}{2\Lambda}\underset{l}{\sum}\{\overline{\nu}_{8}g_{s}G_{\mu\nu}^{\alpha}\sigma^{\mu\nu}(\eta_{L}\nu_{L}+\eta_{R}\nu_{R})+h.c.\}$.

$\,$

Here, $\Lambda$ is compositeness scale, $G_{\mu\nu}^{\alpha}$ is
field strength tensor for gluon, index $\alpha=1,2,..,8$ denotes
the color, $g_{s}$ is the gauge coupling, $\eta_{L}$ and $\eta_{R}$
are the chirality factors, $\nu_{L}$ and $\nu_{R}$ denote left and
right spinor components of neutrino, $\sigma^{\mu\nu}$ is the antisymmetric
tensor. According to neutrino chirality conservation, $\eta_{L}\eta_{R}=0$.
We set $\eta_{L}=1$, $\eta_{R}=0$ in this analysis. In this case
the decay width of $\nu_{8}$ can be written as

$\,$

$\Gamma_{\nu_{8}}=\frac{\alpha_{s}(M_{\nu_{8}})M_{\nu_{8}}^{3}}{4\Lambda^{2}}$.

$\,$

In Fig.2 we present decay width depend on $\nu_{8}$ mass for $\Lambda=10\, TeV$
. Let us mention that decay width is proportional to $\Lambda^{-2}$.

\begin{figure}[H]
\begin{centering}
\includegraphics[scale=0.5]{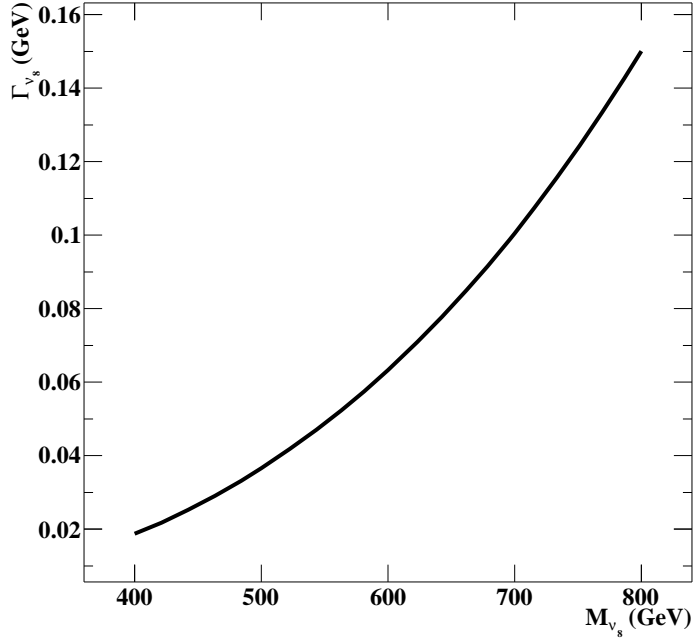}
\par\end{centering}

\caption{Decay width for diffrent $\nu_{8}$ masses.}
\end{figure}

In differ from LQ interpretation {[}2{]} where two decay channels
$\nu_{\tau}+q\rightarrow LQ\rightarrow\tau+q$ and $\nu_{\tau}+q\rightarrow LQ\rightarrow\nu_{\tau}+q$
take place, in $\nu_{8}$ interpretation we deal with $\nu_{8}\rightarrow\nu+g$
only. In {[}2{]} $\nu_{\tau}+q\rightarrow LQ\rightarrow\tau+q$ is
considered as a source fo two $PeV$ energy IceCube events. In this
case approximately $1/6$ of energy is missed due to $\nu_{\tau}$
from $\tau$ decays. For this reason the energy of cosmic neutrinos
is taken between $1-2\, PeV$. In our case half of energy is missed
due to neutrinos from $\nu_{8}$ decays. Therefore we chose $1-4\, PeV$
energy region for cosmic neutrinos resulting in two IceCube $PeV$
events.

In order to perform numerical calculations we implement this lagrangian
into the CalcHEP software \cite{CalcHEP}. In Figure 3 we present
$\nu_{8}$ production cross-section as a function of incoming cosmic
neutrino energy for different $\nu_{8}$ mass values. For numerical
calculations we use $\Lambda=10\, TeV$ together with $CTEQ6L$ parton
distributions. It should be noted that cross-section is proportional
to $\Lambda^{-2}$. 

The event distribution $dN/dE_{\nu}$ is shown in Figure 4. For the
neutrino flux, following \cite{V. Barger}, we use

$\,$

$\Phi_{\nu}^{A-W}=\Phi_{0}\left(\frac{E_{\nu}}{1\, GeV}\right)^{-2.3}$,

$\,$

$\Phi_{0}=6.62\times10^{-7}/GeV/cm^{2}/s/sr$

$\,$

for each neutrino type ($\nu_{e}$, $\nu_{\mu}$ and $\nu_{\tau}$).
Within $LQ$ interpretation only $\nu_{\tau}$ is the source of IceCube
events. In $\nu_{8}$ interpretation there are three possible sources
namely $\nu_{e_{8}}$, $\nu_{\mu_{8}}$and $\nu_{\tau_{8}}$. Real
picture depends on $\nu_{8}$ mass hierarchy. Below we consider a
scenario where one of the color octet neutrinos is lighter than other
two (in degenerate case corresponding cross-sections should be multiplied
by factor 3).

The expected event number at IceCube is estimated using {[}2{]} 

\[
\,
\]

$N=nt\Omega\int dE_{\nu}\sigma_{\nu_{8}}\left|Br\right|\Phi_{\nu}^{A-W}(E_{\nu})$,

\[
\,
\]

where $t=662$ days is the time of exposure, $n=6\times10^{38}$ is
the effective target nucleons number in IceCube and $\Omega=4\pi$.

The IceCube experiment has observed 2 shower events with energies
$1.05\, PeV$ and $1.15\, PeV$ . There are no events between $0.3-1\, PeV$
and above $2\, PeV$. In Table 1 we present the number of events for
different ranges of cosmic ray neutrino energies. The values of $\Lambda$
are scaled in order to handle 2 events in $E_{\nu}=1-4\, PeV$ region
(let us remind that half of energy is missed due to neutrino from
$\nu_{8}$ decay.). For comparison in Table 2 we present the numbers
of $LQ$ events rates from \cite{V. Barger} rescaled in the same
manner. It is seen that both $LQ$ and $\nu_{8}$ interpretations
predict a few events below $1\, PeV$ and most of them lay in $0.3-1\, PeV$
region (see Fig.4 in this paper and Fig.3 from {[}2{]}). Within $\nu_{8}$
interpretation these events can be eliminated increasing $\nu_{8}$
mass, which is not a case for $LQ$ interpretation.

\begin{figure}[H]
\begin{centering}
\includegraphics[scale=0.5]{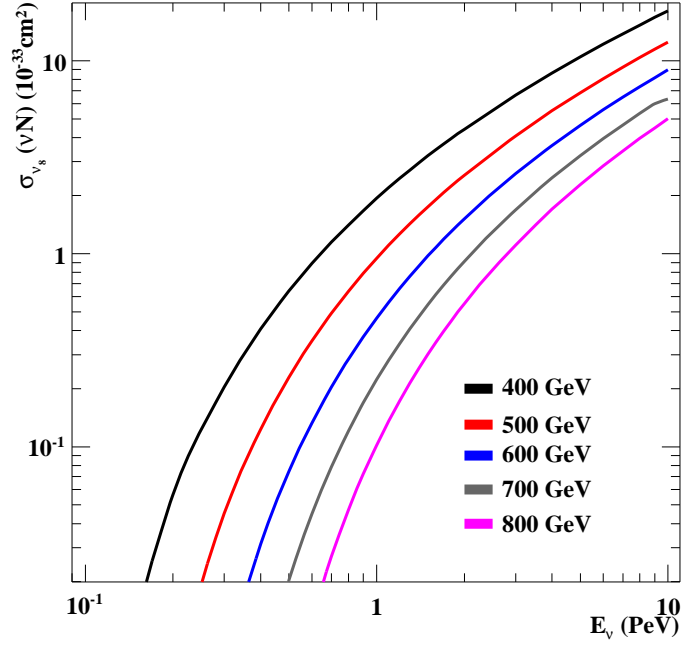}
\par\end{centering}

\caption{$\nu_{8}$ production cross section in $\nu N$ scattering.}
\end{figure}

\begin{figure}[H]
\begin{centering}
\includegraphics[scale=0.5]{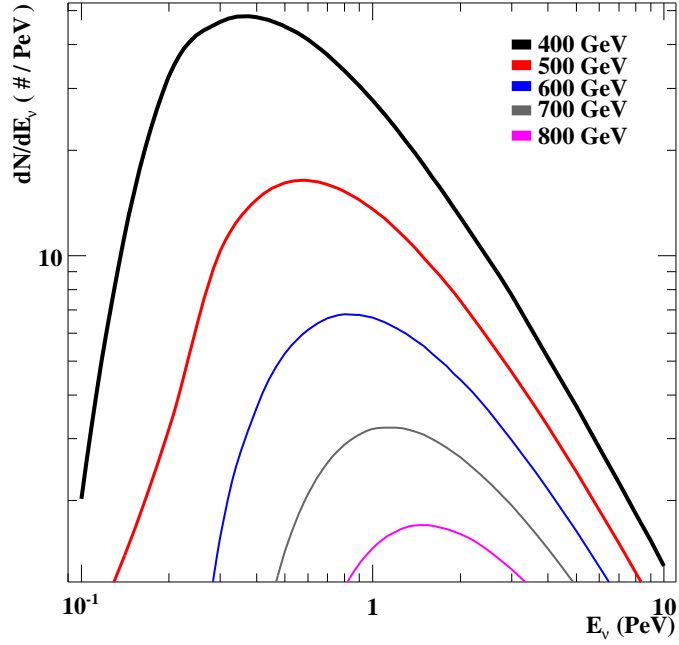}
\par\end{centering}

\caption{Event rate distribution $dN/dE_{\nu}$, from the $\nu_{8}$ cross-section
convoluted with the $A-W$ flux $\Phi_{\nu}^{A-W}$. }
\end{figure}

\begin{table}[H]
\caption{Event numbers for diffent values of $\nu_{8}$ mass and compositeness
scale. }

\centering{}%
\begin{tabular}{|c|c|c|c|c|}
\hline 
$M_{\nu_{8}}(GeV)$  & $\lambda(TeV)$ & $<1\, PeV$  &  $1-4\, PeV$ & $>4\, PeV$ \tabularnewline
\hline 
\hline 
$400$ & $42.1$ & $1.86$ & $2.00$ & $0.86$\tabularnewline
\hline 
$500$  & $31.7$  & $1.12$ & $2.00$ & $1.01$\tabularnewline
\hline 
$600$ & $24.2$ & $0.68$ & $2.00$ & $1.19$\tabularnewline
\hline 
$700$ & $18.7$ & $0.41$ & $2.00$ & $1.41$\tabularnewline
\hline 
$800$ & $14.5$ & $0.24$ & $2.00$ & $1.70$\tabularnewline
\hline 
\end{tabular}
\end{table}

\begin{table}[H]
\caption{Re-scaled event numbers from \cite{V. Barger}.}

\centering{}%
\begin{tabular}{|c|c|c|c|c|}
\hline 
$M_{LQ}(GeV)$ & $f_{L}$  & $<1\, PeV$  & $1-2\, PeV$  & $>2\, PeV$ \tabularnewline
\hline 
\hline 
$500$ & $0.93$ & $7.1$ & $2.0$ & $1.6$\tabularnewline
\hline 
$600$ & $1.2$ & $3.7$ & $2.0$ & $1.6$\tabularnewline
\hline 
\end{tabular}
\end{table}

In conclusion, the IceCube $PeV$ energy neutrino events can be interpreted
as the resonant production of color octet neutrinos with a mass around
$0.5\, TeV$. Correctness of this interpretation may be checked in
near future using forthcoming IceCube and LHC data.

\end{document}